\documentclass[prb,aps,onecolumn,showpacs,floats,a4paper,times,12pt]{revtex4}
\usepackage{epsfig,bm,epsf,graphics}

\begin{document}
\draft
\title{Effects of Ferromagnetic Magnetic Ordering and Phase Transition
on the Resistivity of Spin Current}
\author{K. Akabli and  H. T. Diep\footnote{ Corresponding author, E-mail:
diep@u-cergy.fr}}
\address{Laboratoire de Physique Th\'eorique et Mod\'elisation,
CNRS-Universit\'e de Cergy-Pontoise, UMR 8089\\
2, Avenue Adolphe Chauvin, 95302 Cergy-Pontoise Cedex, France}

\begin{abstract}
It has been shown experimentally a long time ago that the magnetic
ordering causes an anomalous behavior of the electron resistivity in
ferromagnetic crystals. Phenomenological explanations based on the
interaction between itinerant electron spins and lattice spins have
been suggested to explain these observations. We show by extensive
Monte Carlo simulation that this behavior is also observed for the
resistivity of the spin current calculated as a function of
temperature ($T$) from low-$T$ ordered phase to high-$T$
paramagnetic phase in a ferromagnet.  We show in particular that
across the critical region, the spin resistivity undergoes a huge
peak. The origin of this peak is shown to stem from the formation of
magnetic domains near the phase transition.  The behavior of the
resistivity obtained here is compared to experiments and theories. A
good agreement is observed.

\end{abstract}
\maketitle
\section{Introduction}

The interplay between magnetic properties and transport behavior has
been a subject of continuous experimental
investigations.\cite{Shwerer,Stishov,Stishov2} It has been shown in
these works that the resistivity has an anomalous peak in the
magnetic transition temperature region. De Gennes and
Friedel,\cite{DeGennes} and later Fisher and Langer,\cite{Fisher}
have suggested that the peak at the transition is due to the
magnetic short-range interaction.  Various analytical methods such
as mean-field approximations\cite{Kataoka,Wysocki}have been employed
to explain experimental observations.

The problem of spin-dependent transport has been also extensively
studied in magnetic thin films and multilayers.  The so-called
giant magnetoresistance (GMR) was discovered experimentally twenty
years ago.\cite{Baibich,Grunberg} Since then, intensive
investigations, both experimentally and theoretically, have been
carried out.\cite{Fert,review}  Experimental observations show
that when the spin of an itinerant electron is parallel to the
spins of the environment it will go through easily while it will
be stopped if it encounters an antiparallel spin medium.  The
resistance is stronger in the latter case resulting in a GMR.
Although many theoretical investigations have been carried out, to
date very few Monte Carlo (MC) simulations have been performed
regarding the temperature dependence of the dynamics of spins
participating in the current. In a recent work,\cite{Akabli} we
have investigated by MC simulations the effects  of magnetic
ordering on the spin current in magnetic multilayers.  Our results
are in agreement with measurements.\cite{Brucas}

We study  in this paper the transport of itinerant electrons
traveling inside a bulk ferromagnetic crystal using extensive MC
simulations. The electron spin  is supposed to be the Ising model.
Various interactions are taken into account, in particular
interaction between itinerant spins and interaction between
itinerant spins and lattice spins.

The paper is organized as follows. Section II is devoted to the
description of our model and the rules that govern its dynamics.  In
section III, we describe our MC method and discuss the results we
obtained. These results are in agreement with experiments and
theories.  Concluding remarks are given in Section IV.

\section{Model}

We consider in this paper a bulk ferromagnetic crystal. We use the
Ising model and the face-centered cubic (FCC) lattice with size
 $4N_x\times N_y \times N_z$.  Periodic boundary
conditions (PBC) are used in the three directions. Spins localized
at FCC lattice sites are called "lattice spins"
 hereafter.  They interact with each other
through the following Hamiltonian:

\begin{equation}
\mathcal H_l=-J\sum_{\left<i,j\right>}\mathbf S_i\cdot\mathbf S_j,
\label{eqn:hamil1}
\end{equation}
where $\mathbf S_i$ is the Ising spin at lattice site $i$,
$\sum_{\left<i,j\right>}$ indicates the sum over every
nearest-neighbor (NN) spin pair $(\mathbf S_i, \mathbf S_j)$, $J
(>0)$ being the NN interaction.



In order to study the spin transport in the above system, we
consider a flow of itinerant spins interacting with each other and
with the lattice spins.  The interaction between itinerant spins is
defined  as follows,

\begin{equation}
\mathcal H_m=-\sum_{\left<i,j\right>}K_{i,j}\mathbf
s_i\cdot\mathbf s_j,  \label{eqn:hamil2}
\end{equation}
where $\mathbf s_i$ is the Ising spin at position  $\vec r_i$,
and $\sum_{\left<i,j\right>}$ denotes a sum over every spin pair
$(\mathbf s_i, \mathbf s_j)$.  The interaction $K_{i,j}$
depends on the distance between the two spins, i.e.,
$r_{ij}=|\vec r_i-\vec r_j|$.  A specific form of $K_{i,j}$ will
be chosen below.  The interaction between itinerant spins and
lattice spins is given by

\begin{equation}
\mathcal H_r=-\sum_{\left<i,j\right>}I_{i,j}\mathbf
s_i\cdot\mathbf S_j,  \label{eqn:hamil3}
\end{equation}
where the interaction $I_{i,j}$ depends on the distance
between the itinerant spin $\mathbf s_i$ and the lattice spin
$\mathbf S_i$. For the sake of simplicity, we assume the same form
for $K_{i,j}$ and $I_{i,j}$, namely,
$K_{i,j}= K_0\exp(-r_{ij})$ and $I_{i,j}= I_0\exp(-r_{ij})$,
where $K_0$ and $I_0$ are constants.

\begin{figure}
\centerline{\epsfig{file=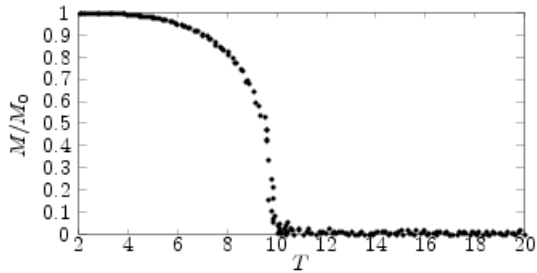,width=2.8in}} \caption{Lattice
magnetization versus temperature $T$. $T_c$ is $\simeq 9.75$ in unit
of $J=1$.} \label{fig:M(T)}
\end{figure}

The procedure used in our simulation is described as follows.
First we study the thermodynamic properties of the bulk system
alone, i.e., without itinerant spins, using Eq. (\ref
{eqn:hamil1}). We perform MC simulations to determine quantities
as the internal energy, the specific heat, layer magnetizations,
the susceptibility, ... as functions of temperature
$T$.\cite{Binder} From these physical quantities we determine the
critical temperature $T_c$ below which the system is in the
ordered phase. We show in Fig. \ref{fig:M(T)} the lattice
magnetization versus $T$.


Once the lattice has been equilibrated at $T$, we inject $N_0$
itinerant spins into the system. The itinerant spins move into the
system at one end, travel in the $x$ direction, escape the system
at the other end to reenter again at the first end under PBC. Note
that PBC are used to ensure that the average density of itinerant
spins remains constant during the time (stationary regime). The
dynamics of itinerant spins is governed by the following
interactions: i) an electric field $\mathbf E$ is applied in the
$x$ direction. Its energy is given by
$\mathcal {H}_E=-\mathbf E \cdot \mathbf v$,
where $ \mathbf v$ is the velocity of the itinerant spin; ii) a
chemical potential term which depends on the concentration of
itinerant spins within a sphere of radius $D_2$ ("concentration
gradient" effect). Its form is given by
$\mathcal {H}_{c}= Dn(\mathbf r)$,
 where $n(\mathbf r)$ is the concentration of
itinerant spins in a sphere of radius $D_2$ centered at $\mathbf r$.
$D$ is a constant taken equal to $K_0$ for simplicity; iii)
interactions between a given itinerant spin and lattice spins inside
a sphere of radius $D_1$ (Eq.~\ref{eqn:hamil3}); iv) interactions
between a given itinerant spin and other itinerant spins inside a
sphere of radius $D_2$ (Eq.~\ref{eqn:hamil2}).

Let us consider the case without an applied magnetic field. The
simulation is carried out as follows: at a given $T$ we calculate
the energy of an itinerant spin by taking into account all the
interactions described above.  Then we tentatively move the spin
under consideration to a new position with a step of length $v_0$ in
an arbitrary direction. Note that this move is immediately rejected
if the new position is inside a sphere of radius $r_0$ centered at a
lattice spin or an itinerant spin. This excluded space emulates the
Pauli exclusion principle in the one hand, and the interaction with
lattice phonons on the other hand.  If the new position does not lie
in a forbidden region of space, then the move is accepted with a
probability given by the standard Metropolis algorithm.\cite{Binder}

\section{Monte Carlo results}

We let $N_0$ itinerant spins travel through the system several
thousands times until a steady state is reached. The parameters we
use in most calculations, except otherwise stated, $s=S=1$ and
$N_x= N_y=N_z= 20$. Other parameters are $D_1=D_2=1$ (in unit of
the FCC cell length), $K_0=I_0=2$, $N_0=20^3$ (namely one
itinerant spin per FCC unit cell), $v_0=1$, $r_0=0.05$. At each
$T$ the equilibration time for the lattice spins lies around
$10^6$ MC steps per spin and we compute statistical averages over
$10^6$ MC steps per spin. Taking $J=1$, we obtain $T_c\simeq 9.75$
for the estimate of the critical temperature of the lattice spins
(see Fig.\ref{fig:M(T)}).

\begin{figure}
\centerline{\epsfig{file=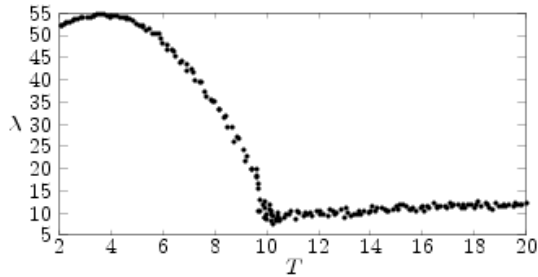,width=2.8in}}
\caption{Characteristic traveling length $\lambda$ in unit of the
FCC cell length versus temperature $T$, for 100 MC
steps.}\label{fig:MFP}
\end{figure}

In Fig. \ref{fig:MFP} we sketch the characteristic traveling
length $\lambda$ computed after a fixed lapse of time as a
function of temperature $T$. As can be seen, $\lambda$ is very
large for $T<T_c$. We note that there is a small depression in the
transition region. We will show below that this has an important
consequence on the spin current: the resistance undergoes a cusp.

\begin{figure}
\centerline{\epsfig{file=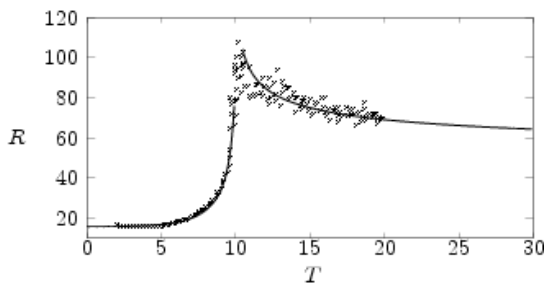,width=2.8in}}
\caption{Resistance $R$ in arbitrary unit versus temperature $T$.
Our result using the Boltzmann equation is shown by the continuous
curve.}\label{resist}
\end{figure}

We define the resistance $R$ as
$R=\frac{1}{n}$,
where $n$ is the number of itinerant spins crossing a unit area
perpendicular to the $x$ direction per unit of time. In
Fig.~\ref{resist} we show the resistance $R$ as a function of
temperature.

There are several striking points observed here. First, $R$ is
very low in the ordered phase and large in the paramagnetic phase.
Below the transition temperature, there exists a single large
cluster with some isolated defects, so that any itinerant spin
having the parallel orientation goes through the lattice without
hindrance. The resistance is thus very small. Secondly, $R$
exhibits a cusp at the transition temperature, the existence of
which is due to the critical fluctuations in the phase transition
region.  De Gennes and Friedel\cite{DeGennes}  and later Fisher
and Langer\cite{Fisher} and then Kataoka\cite{Kataoka} among
others have shown that the resistance is related to the spin
correlation $<S_iS_j>$ which is nothing but the magnetic
susceptibility. The latter diverges at the transition. This
explains at least partially the cusp of $R$ observed here. Another
picture to explain the cusp  is that when $T_c$ is approached
large clusters of up (resp. down) spins form in the critical
region above $T_c$. As a result, the resistance is much larger
than in the ordered phase: itinerant electrons have to steer
around large clusters in order to go through the entire lattice.
Thermal fluctuations are not large enough to allow the itinerant
spin to overcome the energy barrier created by the opposite
orientation of the clusters in this temperature region. Of course,
far above $T_c$, most clusters have a small size, so that the
resistivity is still quite large with respect to the low-$T$
phase. However, $R$ decreases as $T$ is increased because thermal
fluctuations are more and more stronger to help the itinerant spin
overcome energy barriers. Note in addition that the cluster size
is now comparable with the radius $D_1$ of the interaction sphere,
which in turns reduces the height of potential energy barriers. We
have checked this interpretation by first creating an artificial
structure of alternate clusters of opposite spins and then
injecting itinerant spins into the structure.  We observed that
itinerant spins do advance indeed more slowly than in the
completely disordered phase (high-$T$ paramagnetic phase). We have
next calculated directly the cluster-size distribution as a
function of $T$ (not shown) using the Kopelman
algorithm.\cite{Hoshen} The result confirms the effect of clusters
on the spin conductivity.   The reader is referred to our previous
work\cite{Akabli} for results of a multilayer case. Note that in
Fig.~\ref{resist} we also show the results from our theory using
the Boltzmann equation in the relaxation-time approximation.
 The agreement is remarkable.  We will show the details of this theory elsewhere.\cite{Akabli2}


\section{Concluding remarks}
We have shown in this paper first results of MC simulations on the
transport of itinerant spins interacting with localized lattice
spins in a ferromagnetic FCC crystal. Various interactions have been
taken into account. We found that the spin current is strongly
dependent on the lattice spin ordering: at low $T$ itinerant spins
whose direction is parallel (antiparallel) to the lattice spins
yield a strong (weak) current. At high temperatures, the lattice
spins are disordered, the current of itinerant spins is very weak
and does not depend on the input orientation of itinerant spins. As
a consequence, the resistance is very high at high $T$. We found in
the transition region between low-$T$ and high-$T$ phases a peak of
the resistance which is due to the existence of domains of lattice
spins. Experiments and theories support our finding.

The authors are grateful to Sylvain Reynal for discussions.

{}

\end{document}